\documentclass[a4paper,reqno,12pt]{amsart}
\usepackage{amsmath,amsthm,amssymb,amscd,euscript,bbold}
\usepackage{array}
\usepackage[latin1]{inputenc}
\newcommand{\ads}{\mathrm{adS}}
\newcommand{\gS}{\mathfrak{S}}
\newcommand{\gB}{\mathfrak{B}}
\newcommand{\gF}{\mathfrak{F}}

\newcommand{\RR}{\mathbb{R}}
\newcommand{\CC}{\mathbb{C}}
\newcommand{\Spin}{\mathrm{Spin}}
\newcommand{\GL}{\mathrm{GL}}
\DeclareMathOperator{\Cl}{C\ell}
\DeclareMathOperator{\grad}{grad}
\renewcommand{\d}{\partial}
\newcommand{\half}{\tfrac{1}{2}}
\DeclareMathOperator{\laplacian}{\bigtriangleup}
\DeclareMathOperator{\Div}{div}
\newcommand{\eL}{\EuScript{L}}
\newcommand{\M}{\mathsf{M}}
\newcommand{\osp}{\mathfrak{osp}}
\newcommand{\real}[1]{[\![#1]\!]}
\newcommand{\repre}[1]{\underline{\boldsymbol{#1}}}
\newcommand{\so}{\mathfrak{so}}
\newcommand{\su}{\mathfrak{su}}
\renewcommand{\sp}{\mathfrak{sp}}
\renewcommand{\u}{\mathfrak{u}}
\renewcommand{\Sp}{\mathrm{Sp}}

\newcommand{\SU}{\mathrm{SU}}
\newcommand{\SO}{\mathrm{SO}}

\DeclareMathOperator{\pr}{pr}
\newcommand{\1}{\mathbb{1}}
\theoremstyle{plain}

\theoremstyle{definition}

\theoremstyle{remark}

\begin{document}

\title{On the supersymmetries of anti~de~Sitter vacua}
\author{Jos\'e Miguel Figueroa-O'Farrill}
\address{\begin{flushright}Department of Physics\\
Queen Mary and Westfield College\\
Mile End Road\\
London E1 4NS, UK\end{flushright}}
\email{j.m.figueroa@qmw.ac.uk}
\begin{abstract}
  We present details of a geometric method to associate a Lie
  superalgebra with a large class of bosonic supergravity vacua of the 
  type $\ads \times X$, corresponding to elementary branes in
  $\M$-theory and type II string theory.
\end{abstract}
\maketitle

\tableofcontents

\section{Introduction and Summary}

The purpose of the present note is to present the details of a method
to compute the symmetry superalgebra of certain bosonic supergravity
vacua.  This method was applied in \cite{AFHS} on vacua of the form
$\ads \times X$, with $X$ a compact Einstein manifold admitting
Killing spinors, to perform a general geometrical check of the AdS/CFT
duality conjecture \cite{Malda}.  A similar method has been applied by
Gauntlett, Myers and Townsend \cite{GMT1,GMT2,PKT} in the context of
supergravity vacua corresponding to rotating and intersecting branes.
These two methods are conceptually identical, but the computational
details are substantially different to merit the present note.  In
particular, our method is applicable to a large class of examples
where the explicit form of the Killing spinors are not known.
Although most of the results contained in this note were obtained more
than half a year ago, we thought at the time that the method was
well-known.  Only recently have we become aware of the fact that this
might not be the case.

We will be concerned with bosonic vacua of the form $\ads_{p+2} \times
X_d$, where $D=p+2+d$ is either $10$ or $11$, corresponding to type II
and eleven-dimensional supergravities, respectively.  The vacua we
will consider preserve some supersymmetry provided that $X$ admits
real Killing spinors (see below).  In turn this implies that $X$ is a
compact Einstein manifold whose cone has special holonomy.  These
geometries have been classified, and are reviewed in
Table~\ref{tab:baer}.

To every such spacetime $\ads \times X$ we will associate a Lie
superalgebra $\gS$, which is to be understood as the superalgebra of
symmetries of such a background.  We call it the
\emph{symmetry superalgebra} of the bosonic background.

As a vector space, any Lie superalgebra $\gS$ breaks up into an even
and an odd subspace $\gS = \gB \oplus \gF$.  In terms of these
subspaces, the conditions for $\gS$ to be a Lie superalgebra become
the following:
\begin{itemize}
\item[(\textsf{S1})] $\gB$ is a Lie subalgebra;
\item[(\textsf{S2})] $\gF$ is a representation of $\gB$;
\item[(\textsf{S3})] there is a $\gB$-equivariant symmetric bilinear
  map
  \begin{equation}
    \label{eq:FF->B}
    \{-,-\}:\gF \otimes \gF \to \gB
  \end{equation}
  which satisfies the Jacobi identity
  \begin{equation}
    \label{eq:FFFJacobi}
    \{F_1,F_2\} \cdot F_3 + \{F_2,F_3\} \cdot F_1 + \{F_1,F_3\}\cdot
    F_2 = 0~,
  \end{equation}
  where we have denoted the action of $\gB$ on $\gF$ by $\cdot$.
\end{itemize}

The main purpose of this note is to detail the construction of the
symmetry superalgebra of each bosonic vacuum of the form $\ads \times
X$.  The construction will break up into several stages, roughly
corresponding to each of the above structures. The construction runs
as follows:
\begin{itemize}
\item[$\blacktriangleright$] The subspaces $\gB$ and $\gF$ are given
  by the Killing vectors and Killing spinors of the spacetime $\ads
  \times X$;
\item[$\blacktriangleright$] the Killing vectors acts on the Killing
  spinors via the spinorial Lie derivative; and
\item[$\blacktriangleright$] the bilinear map $\gF \otimes \gF \to
  \gB$ is the usual ``squaring'' of spinors.
\end{itemize}

It might seem that in order to identify the symmetry superalgebra of a
given geometry, one needs to know the explicit form of the Killing
spinors.  This is indeed the case in some applications
\cite{GMT1,GMT2,PKT}, but not every time.  If this were the case it
would severely limit the usefulness of this method, since Killing
spinors are not known explicitly for all but the simplest geometries,
namely the space forms.  Fortunately, for the large class of
geometries considered in \cite{AFHS}, one can identify the symmetry
superalgebra using group theory, without the need to construct the
Killing spinors explicitly.  This is made possible by Bär's
observation \cite{Baer} (see also \cite{KlebanovWitten}) that Killing
spinors on $X$ are related to parallel spinors on the cone over $X$,
which have a well-defined group-theoretical interpretation in terms of
the holonomy group of the cone; and by Nahm's classification
\cite{Nahm} of anti~de~Sitter superalgebras, which leaves no room for
ambiguity in the examples we will consider.

The results of \cite{AFHS} are summarised in
Table~\ref{tab:superalgebras}, which may contain some nonstandard
notation.  First of all the notation for Lie superalgebras, although
not traditional, is consistent with the fact that for us $\su_2 \cong
\sp_1$ and \emph{not} $\sp_2$.  Since the spinors in (lorentzian)
eleven and ten dimensions are real, $\gF$ is always a real
representation of $\gB$.  This requires the introduction of some
notation concerning real representations.  Let $R$ be a complex
representation of some given Lie algebra.  There are two ways of
making a real representation out of $R$.  It may be that $R$ has a
real structure (i.e., $R$ admits an invariant symmetric
complex-bilinear form).  In this case, $R$ is the complexification of
a real representation $[R]$.  In other words, $[R]$ is defined by $R
\cong [R]\otimes_\RR \CC$.  On the other hand, if $R$ is truly a
complex representation, we can consider $R\oplus R^*$.  This
representation has a real structure, so that $R\oplus R^* \cong
[R\oplus R^*] \otimes_\RR \CC$.  The real representation $[R\oplus
R^*]$ is denoted $\real{R}$.  Notice that if $R$ has complex dimension
$r$, then $[R]$ has real dimension $r$ and $\real{R}$ has real
dimension $2r$.  If $R$ and $R'$ admit quaternionic structures (i.e.,
they admit invariant antisymmetric complex-bilinear forms), then their
tensor product $R\otimes_\CC R'$ inherits a real structure and it
makes sense to consider the real representation $[R\otimes_\CC R']$.
This explains the first fermionic representation in the Table.

\begin{table}[h!]
\centering
\setlength{\extrarowheight}{3pt}
\begin{tabular}{|>{$}c<{$}|c|>{$}c<{$}|>{$}c<{$}|>{$}c<{$}|}\hline
(p,d) & $X_d$ & \gB & \gF & \gS\\
\hline
\hline
(5,4) & sphere & \so_{6,2} \times \sp_2 & [(\repre{8},\repre{4})] &
\osp_{6,2|2}\\
(3,5) & sphere & \so_{4,2} \times \su_4 & \real{(\repre{4},\repre{4})} 
& \su_{2,2|4}\\
(3,5) & Sasaki--Einstein & \so_{4,2} \times \u_1 &
\real{(\repre{4},\repre{1})} & \su_{2,2|1}\\ 
(2,7) & sphere & \so_{3,2} \times \so_8 & (\repre{4},\repre{8}) &
\osp_{8|2}\\
(2,7) & 3-Sasaki & \so_{3,2} \times \so_3 & (\repre{4},\repre{3}) &
\osp_{3|2}\\
(2,7) & Sasaki--Einstein & \so_{3,2} \times \so_2 &
(\repre{4},\repre{2}) & \osp_{2|2}\\
(2,7) & nearly-parallel $G_2$ & \so_{3,2} \times \so_1 & \repre{4} &
\osp_{1|2}\\[3pt]
\hline
\end{tabular}
\vspace{8pt}
\caption{Symmetry superalgebras of $\ads_{p+2} \times X_d$, with $X$
  simply-connected.}
\label{tab:superalgebras}
\end{table}

The Table is complete for simply-connected spaces $X$.  It is possible
to obtain other symmetry superalgebras by taking finite quotients of
the sphere.  The main purpose of the note is to exhibit the
computations leading to this Table in some detail. 

This note is organised as follows.  In Section~2 we define real
Killing spinors and introduce the main ingredients in the construction
of the isometry subalgebra: the spinorial Lie derivative and the
squaring of the spinors.  In Section~3 we will discuss the main
computational tool in our approach: the relation between Killing
spinors on $X$ and parallel spinors on the cone over $X$.  In
Section~4 we discuss the relation between the isometries of $X$ and
the isometries of the cone.  Finally in Section~5 we will apply this
to the cases of interest $\ads_{p+2} \times X_d$ for
$(p,d)\in\{(2,7),(3,5),(5,4)\}$.

\section{Killing spinors and the symmetry superalgebra}

In this section we introduce the basic ingredients in the construction 
of the symmetry superalgebra of a bosonic vacuum of the form $\ads
\times X$.  Such vacua will preserve supersymmetry provided that $X$
admits real Killing spinors.  Let us review this notion.

Let $X$ be an $n$-dimensional riemannian spin manifold.  Let $\Psi$ be
a spinor on $X$.  We say that $\Psi$ is a \emph{real Killing spinor}
if
\begin{equation}
  \label{eq:KillingSpinor}
  \nabla_W \Psi = \lambda W \cdot \Psi\quad\text{for all vectors $W$,} 
\end{equation}
where $\lambda \in \RR$ is a constant, and $\cdot$ is the action of
the Clifford bundle on the spinor bundle.  Relative to a local
orthonormal frame, this equation becomes:
\begin{equation}
  \label{eq:KillingSpinorFrame}
    \nabla_i \Psi = \lambda \gamma_i\cdot\Psi~,
\end{equation}
where $\gamma_i$ satisfy $\{\gamma_i,\gamma_j\} = -2\,\delta_{ij}$.
It should be remarked that the ``real'' in the definition refers to
the fact that $\lambda$ is real: the spinor itself is complex.  The
integrability condition for \eqref{eq:KillingSpinor} says that $X$ is
an Einstein manifold with scalar curvature $4\,\lambda^2 n(n-1)$.
Such manifolds are necessarily compact.  We choose to normalise the
metric in such a way that $\lambda = \pm \half$.

Of the two possible values of $\lambda$, only one will give rise to
symmetries of the supergravity vacuum; indeed, the sign of $\lambda$
is correlated to the sign of the flux through $X_d$ of the $d$-form
field in the supergravity theory under consideration.

The isometry algebra $\gB$ of $\ads_{p+2} \times X_d$ is isomorphic to
the product $\gB(\ads_{p+2})\times \gB(X_d)$ of the isometry algebras
of the anti~de~Sitter space $\ads_{p+2}$ and of $X_d$.  The fermionic
subspace $\gF$ will similarly break up into a direct sum
$\gF(\ads_{p+2}) \oplus \gF(X_d)$ of the spaces of Killing spinors on
$\ads_{p+2}$ and on $X_d$.  The isometries and Killing spinors on
anti~de~Sitter space are well known and will be discussed briefly
below.  On the other hand, the Killing spinors on $X$ can be mapped to
parallel spinors on the cone over $X$, and can be studied
group-theoretically.  This will be discussed in detail below.

\subsection{The spinorial Lie derivative}

The next ingredient in the construction of the symmetry superalgebra
is the spinorial Lie derivative, which tells us how the Killing
vectors $\gB$ act on the Killing spinors $\gF$.  Since the spinor
bundle is not a $\GL_n$ bundle, a Lie derivative cannot be readily
defined.  However we will see that for certain types of vector fields,
namely the (conformal) Killing vectors, we will be able to make sense
of the Lie derivative of a spinor.

Let $V$ be a vector field and let $\Psi$ be a spinor. The spinorial
Lie derivative $\eL_V$ must obey the following properties:
\begin{itemize}
\item[(\textsf{L1})] it should be a derivative; that is, for any
function $f$,
  \begin{equation*}
    \eL_V (f\Psi) = V(f)\, \Psi + f \eL_V \Psi~;
  \end{equation*}
\item[(\textsf{L2})] it should be independent on the choice of local
orthonormal frame;
\item[(\textsf{L3})] it should induce, on bispinors, the usual Lie
derivative on differential forms; and
\item[(\textsf{L4})] it should form a representation of (a Lie
  subalgebra of) the algebra of vector fields:
  \begin{equation*}
    [\eL_V,\eL_W]\, \Psi = \eL_{[V,W]}\, \Psi~.   
  \end{equation*}
\end{itemize}

Let us see what it takes to satisfy these conditions.  The first
condition (\textsf{L1}) simply says that
\begin{equation*}
  \eL_V \Psi = \nabla_V \Psi + \theta(V)\cdot \Psi~,
\end{equation*}
where $\nabla$ is the spin connection and $\theta(V)$ takes values in
the Clifford bundle $\Cl(TX)$, which we identify with the
endomorphisms of the spinor bundle.  This expression also satisfies
the second condition (\textsf{L2}), since both the spin connection
$\nabla$ and the sections of the Clifford bundle transform covariantly
under a change of local orthonormal frame.

The third condition (\textsf{L3}) is tantamount to imposing that the
spinorial Lie derivative be compatible with the action of the Clifford
bundle:
\begin{equation}
  \label{eq:CliffordBundle}
  \eL_V W\cdot \Psi = [V,W]\cdot \Psi + W \cdot \eL_V \Psi~,
\end{equation}
which in turn implies the following relation
\begin{equation*}
  \theta(V)\cdot W - W \cdot \theta(V) = - \nabla_W V\quad\text{for
  all $W$,}
\end{equation*}
which is to be understood as a relation in the Clifford bundle.  Up to
central terms in the Clifford algebra, $\theta(V)$ must take the form
\begin{equation*}
  \theta(V) = \tfrac{1}{4} \theta_{ij} \gamma^{ij}~,
\end{equation*}
where the coefficients $\theta_{ij} = - \theta_{ji}$ must satisfy the
following condition
\begin{equation*}
  \theta_{ij} = - \nabla_i V_j~.
\end{equation*}
Taking into account that the left-hand side is antisymmetric, the
only solution is $\theta_{ij} = - \nabla_{[i}V_{j]}$, where
$\nabla_{(i} V_{j)}=0$.  In other words, $V$ has to be a Killing
vector.

In summary, we define the \emph{Lie derivative} of a spinor $\Psi$ in
the direction of the Killing vector $V$ by
\begin{equation}
  \label{eq:LieDerivative}
  \eL_V \Psi \equiv \nabla_V \Psi - \tfrac{1}{4} (\nabla_i V_j)
  \gamma^{ij} \cdot \Psi~.
\end{equation}
This equation is due to Kosmann \cite{Kosmann}, where it is shown that 
the fourth condition (\textsf{L4}) is automatically satisfied.

It should be remarked that if we drop property \textsf{L3} then the
spinorial Lie derivative \eqref{eq:LieDerivative} obeys property
\textsf{L4} provided that $V$ is a \emph{conformal} Killing
vector (see, e.g., \cite{Spindel}).

If $V$ is a Killing vector the spinorial Lie derivative preserves the
spin connection
\begin{equation}
  \label{eq:LDKillingVector}
  \eL_V \nabla_W \Psi = \nabla_{[V,W]} \Psi + \nabla_W \eL_V \Psi~.
\end{equation}
It follows from this fact that the spinorial Lie derivative with
respect to a Killing vector preserves the space of Killing spinors
(see, e.g., \cite{Moroianu}).  Indeed, suppose that $\Psi$ is a
Killing spinor, and let us take the Lie derivative of equation
\eqref{eq:KillingSpinor}:
\begin{multline*}
  \eL_V \left[ \nabla_W \Psi - \lambda W \cdot \Psi \right]\\
  = \nabla_{[V,W]} \Psi + \nabla_W \eL_V \Psi - \lambda [V,W]\cdot \Psi
  - \lambda W \cdot \eL_V \Psi~,
\end{multline*}
where we have used equation \eqref{eq:LDKillingVector}.  The first and
third terms in the right-hand side cancel each other out because
$\Psi$ is a Killing spinor, as does the left-hand side of the
equation. The remaining terms say that $\eL_V \Psi$ is a Killing
spinor with the same constant $\lambda$:
\begin{equation*}
  \nabla_W \eL_V \Psi = \lambda W \cdot \eL_V \Psi~.
\end{equation*}

Although we have been discussing the case of $\lambda\neq0$, the above 
discussion applies equally well to manifolds admitting parallel
spinors, which we can think of as Killing spinors with $\lambda=0$.

\subsection{From Killing spinors to Killing vectors}

The final ingredient in the construction of the symmetry superalgebra
is the bilinear map taking Killing spinors to Killing vectors.  Let
$\Psi$ and $\Xi$ be two real Killing spinors with the same constant
$\lambda$.  Then we can define a vector $V$ whose components relative
to a local orthonormal frame $\{E_i\}$ are given by
\begin{equation}
  \label{eq:Vector}
  V^i = \langle \Psi, \gamma^i \cdot \Xi\rangle~,
\end{equation}
where $\langle-,-\rangle$ is the hermitian inner product in the spinor 
representation.  We will show that $V$ is a Killing vector.

To see this we notice that for the Clifford algebra $\Cl_n$ given by
\eqref{eq:Cl(n)}, we can always choose the inner product
$\langle-,-\rangle$ such that the Clifford action is unitary:
\begin{equation}
  \label{eq:InnerProduct}
  \langle \gamma^i \cdot \Psi, \Xi \rangle = - \langle \Psi,\gamma^i
  \cdot \Xi\rangle~.
\end{equation}

We can now simply compute:
\begin{align*}
  \nabla_i V_j
  &= \langle \nabla_i \Psi, \gamma_j \cdot \Xi\rangle +
  \langle \Psi, \gamma_j \cdot \nabla_i \Xi\rangle\\
  &= \lambda \langle \gamma_i \Psi, \gamma_j \cdot \Xi\rangle +
  \lambda \langle \Psi, \gamma_j \gamma_i \cdot \Xi\rangle\\
  &= \lambda \langle \Psi, [\gamma_j, \gamma_i] \cdot \Xi \rangle~,
\end{align*}
which is antisymmetric under $i\leftrightarrow j$.  Thus Killing's
equation is satisfied.

It might seem that equation \eqref{eq:InnerProduct} makes the mapping
$\gF \otimes \gF \to \gB$ antisymmetric instead of symmetric.  This is
only because geometric spinors are commuting.  On the other hand, the
field-theoretical spinors (hence the objects in $\gF$) are
anticommuting: they can be thought of as products of the Killing
spinors with anticommuting elements of an underlying infinitely
generated Grassmann algebra.  In that case, we the map $\gF \otimes
\gF \to \gB$ is indeed symmetric as expected.  Moreover because of
property \textsf{L3} of the spinorial Lie derivative, it is
$\gB$-equivariant.

The final property that has to be checked is the Jacobi identity for
the trilinear map $\gF \otimes \gF \otimes \gF \to \gF$.  If $\Psi_a$
for $a=1,2,3$ are Killing spinors with the same constant $\lambda$,
then the Jacobi identity becomes
\begin{equation}
  \label{eq:Jacobi}
  \eL_{V_{12}} \Psi_3 + \eL_{V_{23}} \Psi_1 + \eL_{V_{31}} \Psi_2 = 0~,
\end{equation}
where $V_{ab}$ is the Killing vector made out of $\Psi_a$ and
$\Psi_b$, whose components are given by $V_{ab}^i = \langle \Psi_a,
\gamma^i \cdot \Psi_b\rangle$.  Applying the definitions, the Jacobi
identity \eqref{eq:Jacobi} becomes
\begin{equation}
  \label{eq:JacobiToo}
    \langle \Psi_1, \gamma^i \cdot \Psi_2 \rangle \, \gamma_i \cdot
    \Psi_3 + \half \langle \Psi_1, \gamma^{ij} \cdot \Psi_2 \rangle \, 
    \gamma_{ij} \cdot \Psi_3 + \text{cyclic} = 0~,
\end{equation}
which, if need be, can be checked case by case using Fierz
rearrangements.

\section{Killing spinors and parallel spinors}

In this and the following section we collect some necessary
geometrical facts that will allow us to turn the determination of the
symmetry superalgebra $\gS$ into a group theory problem, at least for
the cases we will consider.  The observation underpinning this
approach is due to Bär \cite{Baer}, who noticed that the Killing
spinor equation \eqref{eq:KillingSpinor} normalised to $\lambda=\pm
\half$ can be understood as the condition that $\Psi$ be parallel
with respect to a modified connection $\Tilde\nabla$ which coincides
formally with the riemannian connection on the metric cone of $X$.  In 
fact, this can be made precise and Bär proved that there is a
one-to-one correspondence between Killing spinors on $X$ and parallel
spinors on the cone.

The cone $\Tilde X$ of $X$ is topologically $\RR_+ \times X$ with
metric
\begin{equation}
  \label{eq:ConeMetric}
  \Tilde g = dr^2 + r^2 g~,
\end{equation}
where $g$ is the metric on $X$ and $r>0$ parametrises $\RR_+$.  The
manifold $X$ is isometric to the $r=1$ slice of $\Tilde X$, and we
shall not distinguish between them.  For $X$ Einstein with scalar
curvature $n(n-1)$, (where $n= \dim X$), the cone metric is
Ricci-flat.

Let $\xi = r\d_r$ be the \emph{Euler vector} on $\Tilde X$; it
generates an infinitesimal homothety.  Any vector field $V$ on $X$ can
be lifted to a unique vector field on $\Tilde X$ orthogonal to $\xi$
and such that it projects to $V$ under the natural projection $\Tilde
X = \RR_+ \times X \to X$.  We shall not distinguish between a vector
and its lift.  Notice however that if $V,W$ are vector fields on $X$,
we have that
\begin{equation*}
  \Tilde g (V,W) = r^2 \, g(V,W)~.
\end{equation*}
Let $\Tilde\nabla$ denote the riemannian connection on the cone
$\Tilde X$.  Then we have the following:
\begin{equation}
  \label{eq:ConeConnection}
  \Tilde\nabla_\xi \xi = \xi~,\quad
  \Tilde\nabla_\xi V = V\quad\text{and}\quad
  \Tilde\nabla_V W = \nabla_V W - g(V,W) \xi~,
\end{equation}
for all vectors $V,W$ tangent to $X$.  In fact, Gibbons and Rychenkova
\cite{GR2} have proven that the characterising property of a metric
cone is the existence of a vector field $\xi$ such that
$\Tilde\nabla_\xi V = V$ for \emph{all} vector fields $V$.

Let $\{E_i\}$ be a local orthonormal frame for $X$, and $\{\Tilde
E_I\} = \{\Tilde E_i \equiv \frac{1}{r} E_i, \Tilde E_r \equiv \d_r\}$
the induced local orthonormal frame for $\Tilde X$.  Define
$\Tilde\omega_{I}{}^J{}_K$ by
\begin{equation}
  \label{eq:SpinConnection}
  \Tilde\nabla_{\Tilde E_I} \Tilde E_J = \Tilde\omega_{I}{}^J{}_K \,
  \Tilde E_K~.
\end{equation}
A quick calculation shows that
\begin{equation}
  \label{eq:SpinConnectionCone}
  \Tilde\omega_{r}{}^J{}_K = 0~,\quad
  \Tilde\omega_{i}{}^j{}_r = \frac{1}{r} \delta_i{}^j~,\quad
  \Tilde\omega_{i}{}^r{}_j = - \frac{1}{r} g_{ij}\quad\text{and}\quad
  \Tilde\omega_{i}{}^j{}_k = \frac{1}{r} \omega_{i}{}^j{}_k~.
\end{equation}

Now suppose that $\Psi$ is a parallel spinor on $\Tilde X$:
\begin{equation}
  \label{eq:ParallelSpinor}
  \Tilde\nabla_{\Tilde E_I} \Psi = \Tilde E_I^\mu \d_\mu \Psi -
  \tfrac{1}{4} \Tilde\omega_I{}^{JK} \Gamma_{JK}\cdot \Psi = 0~.
\end{equation}
In terms of the explicit expression \eqref{eq:SpinConnectionCone}, we
have that
\begin{equation}
  \label{eq:ParallelSpinorCone}
  \Tilde\nabla_{\Tilde E_r} \Psi = \d_r \Psi = 0\quad\text{and}\quad
  \Tilde\nabla_{\Tilde E_i} \Psi = \frac{1}{r} \left( \nabla_{E_i}
  \Psi - \half \Gamma_i \Gamma_r \cdot \Psi \right) = 0~.
\end{equation}

In order to relate this equation to the Killing equation
\eqref{eq:KillingSpinorFrame} on $X$ we need to recall how the
Clifford bundles on $\Tilde X$ and on $X$ are related.

Let $\Cl(TX)$ be the Clifford bundle on $X$, which is a bundle of
Clifford algebras isomorphic to $\Cl_n$, the euclidean Clifford
algebra in $n$-dimensions.  $\Cl_n$ is generated by $\{\gamma_i\}$
subject to
\begin{equation}
  \label{eq:Cl(n)}
  \{\gamma_i,\gamma_j\} = -2\, \delta_{ij}~.
\end{equation}
On the other hand, the Clifford bundle $\Cl(T\Tilde X)$ on the cone is 
locally modelled on $\Cl_{n+1}$, which is generated by $\{\Gamma_I\}$
with $I=(i,r\equiv n+1)$ with $i$ running from $1$ to $n$, subject to
\begin{equation}
  \label{eq:Cl(n+1)}
  \{\Gamma_I,\Gamma_J\} = -2\, \delta_{IJ}~.
\end{equation}
The algebra $\Cl_n$ is naturally a subalgebra of $\Cl_{n+1}^0$, the
even subalgebra of $\Cl_{n+1}$.  The embedding is given by
\begin{equation}
  \label{eq:Cl(n)->Cl(n+1)}
  \gamma_i \mapsto \varepsilon \Gamma_i \Gamma_r~,
\end{equation}
where $\varepsilon^2=1$.  Notice that under this map $\half
\gamma_{ij} \mapsto \half \Gamma_{ij}$, so that it induces the
natural embedding of the Spin groups.

Using this we can rewrite equation \eqref{eq:ParallelSpinorCone} as
\begin{equation}
  \label{eq:KillingSpinorTwo}
  \d_r \Psi = 0 \quad\text{and}\quad
  \nabla_i \Psi = \half \varepsilon \gamma_i \cdot \Psi~.
\end{equation}
Therefore we deduce that there is a one-to-one correspondence between
Killing spinors on $X$ and parallel spinors on the cone $\Tilde X$: a
parallel spinor $\Psi$ on the cone restricts (at $r=1$) to a Killing
spinor on $X$, and conversely, given a Killing spinor on $X$ we can
extend this to a parallel spinor on the cone by demanding that it does
not depend on $r$.

In order to understand what kind of Killing spinors we get (i.e., the
sign of $\varepsilon$) we need to look more closely at the embedding
\eqref{eq:Cl(n)->Cl(n+1)}.  Notice that it has the following
additional property:
\begin{equation}
  \label{eq:VolumeElement}
  \gamma_1 \cdots \gamma_n \mapsto
  \begin{cases}
    \varepsilon \Gamma_1 \cdots \Gamma_{n+1}~,& \text{if $n$ is odd;
    and}\\
    \Gamma_1 \cdots \Gamma_n~, & \text{if $n$ is even.}
  \end{cases}
\end{equation}
Let $n$ be odd.  In an irreducible representations of $\Cl_n$, the
volume element $\gamma_1 \cdots \gamma_n$ is a scalar multiple of the
identity.  According to the above equation, it gets mapped to
$\varepsilon$ times the volume element of $\Cl_{n+1}$.  This means that
$\varepsilon$ is fixed in terms of the chirality of the spinor $\Psi$: 
the nature of the correspondence will depend on which irreducible
representation we have chosen to work with in $\Cl_n$---equivalently, 
the orientation of $X$.  On the other hand, if $n$ is even,
$\varepsilon$ is not fixed.  Therefore for each parallel spinor on
$\Tilde X$, we get one parallel spinor with $\varepsilon = 1$ and one
with $\varepsilon = -1$ simply by choosing one of the two inequivalent 
irreducible representations of $\Cl_{n+1}$.

Since $X$ admits real Killing spinors, it is Einstein with positive
scalar curvature.  By Myer's theorem (see, e.g., \cite{CheegerEbin})
it is compact, and hence its fundamental group is finite.  We will
moreover assume that $X$ is simply connected.  This allows us to use a
result of Gallot's \cite{Gallot} quoted in \cite{Baer}, which says
that the cone over a compact simply-connected manifold is either flat,
so that the manifold is the round sphere, or irreducible.  Finally,
the simply-connected irreducible manifolds admitting parallel spinors
have been classified by Wang \cite{Wang}.  The result is summarised in
Table~\ref{tab:wang}.  The last column indicates the number $N$ of
parallel spinors, which in even dimensions has been refined according
to chirality.

\begin{table}[h!]
\centering
\setlength{\extrarowheight}{3pt}
\begin{tabular}{|>{$}c<{$}|>{$}c<{$}|c|>{$}c<{$}|}\hline
\dim & \text{Holonomy} & Geometry& N\\[3pt]
\hline
\hline
4k & \Sp_k & hyperk\"ahler & (k+1,0)\\
4k & \SU_{2k} & Calabi--Yau & (2,0)\\
4k+2 & \SU_{2k+1} & Calabi--Yau & (1,1)\\
7 & G_2 & parallel $G_2$ & 1\\
8 & \Spin_7  & parallel $\Spin_7$ & (1,0)\\[3pt]
\hline
\end{tabular}
\vspace{8pt}
\caption{Simply-connected irreducible manifolds admitting parallel
spinors}
\label{tab:wang}
\end{table}

We can then finally write down the possible geometries of
simply-connected Einstein manifolds admitting $N_\pm$ real Killing
spinors with $\lambda = \pm \half$.  This is summarised in
Table~\ref{tab:baer}.

\begin{table}[h!]
\centering
\setlength{\extrarowheight}{3pt}
\begin{tabular}{|>{$}c<{$}|c|c|>{$}c<{$}|}\hline
\dim X & Geometry & Cone & (N_+, N_-)\\
\hline
\hline
d & round sphere & flat & (2^{\lfloor d/2\rfloor},2^{\lfloor
d/2\rfloor})\\
4k-1 & $3$-Sasaki & hyperkähler & (k+1,0)\\
4k-1 &Sasaki--Einstein & Calabi--Yau & (2,0)\\
4k+1 &Sasaki--Einstein & Calabi--Yau & (1,1)\\
6 & nearly K\"ahler & parallel $G_2$ & (1,1)\\
7 & nearly parallel $G_2$ & parallel $\Spin_7$ & (1,0)\\[3pt]
\hline
\end{tabular}
\vspace{8pt}
\caption{Simply-connected manifolds admitting real Killing spinors}
\label{tab:baer}
\end{table}

We can summarise this section conceptually as follows.  Let
$\gF_\pm(X)$ denote the space of Killing spinors on $X$ with
$\lambda=\pm\half$ and let $\gF_\pm(\Tilde X)$ (or simply $\gF(\Tilde
X)$ if $\dim \Tilde X$ is odd) be the space of parallel spinors (of
definite chirality, if applicable) on $\Tilde X$.  Then there are
isomorphisms $\gF_\pm (X) \cong \gF_\pm(\Tilde X)$, for $\dim\Tilde X$
even, and $\gF_\pm (X) \cong \gF(\Tilde X)$ for $\dim \Tilde X$ odd.
We will simply summarise this family of isomorphisms as $\gF(X) \cong
\gF(\Tilde X)$.

\section{Isometries of a conical geometry}

In this section we will characterise the isometries of the cone
metric $\Tilde g$ on $\Tilde X$ in terms of data on the original space 
$X$.  We will see that they come in two flavours: either they are lifts
of Killing vectors on $X$, or they are related to conformal Killing
vectors on $X$ which are given by gradients of eigenfunctions of the
Laplacian on $X$.  These latter Killing vectors only exist when $X$ is 
a spherical form.

Let $\Tilde V$ be a Killing vector on $\Tilde X$, so it satisfies
Killing's equation
\begin{equation}
  \label{eq:KillingVector}
  \Tilde g (\Tilde\nabla_U \Tilde V, W) + 
  \Tilde g (U, \Tilde\nabla_W \Tilde V) = 0~,
\end{equation}
for all vector fields $U,W$ in $\Tilde X$.  Let us write $\Tilde V = f
\, \d_r + \Bar V$, where $f$ is a function on $\Tilde X$ and $\Bar V$
is a vector field on $\Tilde X$ orthogonal to $\d_r$.  It is a simple
matter to compute $\Tilde\nabla \Tilde V$ using equation
\eqref{eq:ConeConnection}:
\begin{align*}
  \Tilde\nabla_r \Tilde V &= \d_r f\, \d_r + \d_r \Bar V + \frac{1}{r}
  \Bar V\\
  \Tilde\nabla_W \Tilde V &= Wf\, \d_r + \frac{1}{r} f W +
  \nabla_W \Bar V - r \, g(W,\Bar V) \, \d_r~,
\end{align*}
where $W$ is tangent to $X$.  Therefore $\Tilde V$ satisfies Killing's
equation \eqref{eq:KillingVector} if and only if $f$ and $\Bar V$
satisfy:
\begin{gather*}
  \d_r f= 0~,\quad
  Wf = - r^2 \, g(\Bar V,W)\quad\text{and}\\
  g(\nabla_W \Bar V, U) + g ( W, \nabla_U \Bar V) = - \frac{2}{r} f\,
  g(U,W)~,
\end{gather*}
for every $U,W$ tangent to $X$.  The first equation says that $f$ is
the lift of a function on $X$.  The second equation says that
\begin{equation*}
  \Bar V = V + \frac{1}{r} \grad f~,
\end{equation*}
where $\d_r V = 0$, so that $V$ is the lift of a vector field on $X$.
The third (and last) equation becomes:
\begin{multline*}
  g(\nabla_W V, U) + g(W, \nabla_U V)\\
  + \frac{1}{r}\left[ g\left(\nabla_W \grad f, U\right) + 
    g\left(W, \nabla_U \grad f\right) + 2 f\, g(U,W)\right] = 0~.
\end{multline*}
Since the first two terms are independent of $r$ and so is the term
inside the square brackets, we see that they both have to vanish
separately.  This means that $V$ is a Killing vector on $X$ and that
$f$ is a function such that its gradient is a conformal Killing
vector:
\begin{equation*}
  g\left(\nabla_W \grad f, U\right) +  g\left(W, \nabla_U \grad
  f\right) = - 2 f\, g(U,W)~.
\end{equation*}
Tracing this equation, we see that $f$ must in addition satisfy
\begin{equation}
  \label{eq:Obata}
  \laplacian f = - \Div\grad f = n\,f~,\quad\text{with $n=\dim X$.}
\end{equation}
For $X$ an Einstein manifold, this equation is equivalent to the
celebrated \emph{Obata equation} \cite{Obata}, which only has
solutions if $X$ is locally isometric to a sphere.  In other words,
if $X$ is simply connected, then $X\cong S^n$ and there are $n+1$
functions obeying \eqref{eq:Obata}: the first nonconstant spherical
harmonics, which transform according to the vector representation of
$\SO_{n+1}$.  If $X$ is not simply connected, then $X \cong
S^n/\Gamma$, where $\Gamma \subset \SO_{n+1}$ is a finite subgroup.
The number of solutions of \eqref{eq:Obata} will then be equal to the
dimension of the space of $\Gamma$-invariant solutions of the equation 
on the sphere; in other words, the number of linearly independent
singlets in the decomposition of the vector representation of
$\SO_{n+1}$ under the subgroup $\Gamma$.

In summary, the Killing vectors of a cone metric $(\Tilde X, \Tilde
g)$ are of two types:
\begin{itemize}
\item lifts of Killing vectors of $(X,g)$; and
\item vectors of the form $f \d_r + \frac{1}{r} \grad f$, where $f$ is
      a function on $X$ obeying $\laplacian f = n\,f$;
\end{itemize}
with the latter only existing in the case of spherical forms.  We can
reformulate the former point more precisely as follows.

Let $\gB(X)$ and $\gB(\Tilde X)$ denote the Lie algebras of Killing
vectors on $X$ and $\Tilde X$, respectively.  Then lifting the Killing
vectors on $X$ to $\Tilde X$ gives rise to a Lie algebra homomorphism
$\gB(X) \to \gB(\Tilde X)$.  Because lifting at $r=1$ is an isometry,
this homomorphism is actually one-to-one. To prove that the lift is a
Lie algebra homomorphism we need to go into a little bit more detail.

\begin{description}
\small
\item[\textbf{Proof}]\hspace*{\fill}\linebreak[4]
  We require a more precise definition of the lift to $\Tilde X$ of a
  vector field on $X$.  As smooth manifolds, Ignoring the metric,
  $\Tilde X = \RR_+ \times X$.  Let $\pr_1: \Tilde X \to \RR_+$ and
  $\pr_2: \Tilde X \to X$ be the canonical projections.  Let $V$ be a
  vector field on $X$.  Its lift to $\Tilde X$ is the unique vector
  field $\Tilde V$ obeying $(\pr_1)_* \Tilde V = 0$ and $(\pr_2)_*
  \Tilde V = V$.  Now let $W$ be another vector field on $X$ and
  $\Tilde W$ its lift to $\Tilde X$.  Then by the functoriality of the
  derivative map
  \begin{align*}
    &(\pr_1)_* [\Tilde V, \Tilde W] = [(\pr_1)_* \Tilde V,
    (\pr_1)_*\Tilde W] = 0~,\text{and}\\
    &(\pr_2)_* [\Tilde V, \Tilde W] = [(\pr_2)_* \Tilde V,
    (\pr_2)_*\Tilde W] = [V,W]~,
  \end{align*}
  whence $[\Tilde V, \Tilde W]$ is the lift of $[V,W]$.
\end{description}

\section{Anti~de~Sitter supersymmetries}

We have now all the ingredients necessary to determine the symmetry
superalgebra of a bosonic background of the form $\ads_{p+2} \times
X_d$ for $(p,d)\in\{(2,7),(3,5),(5,4)\}$.  We will follow the
following strategy:
\begin{enumerate}
\item we determine the isometry algebra $\gB$ of the geometry $\ads
  \times X$;
\item we determine the representation $\gF$ of $\gB$ under which the
  Killing spinors transform; and
\item we inspect Nahm's classification for candidate Lie
  superalgebras.
\end{enumerate}
The second step makes use of the results about conical geometry
developed in the previous two sections, and turns the problem into a
group-theoretical one.

We saw in the previous two sections that there is a Lie algebra
homomorphism $\gB(X) \to \gB(\Tilde X)$ and a vector space isomorphism
$\gF(X) \cong \gF(\Tilde X)$.  Moreover, as we now show, these maps
make the following diagram commute, where the horizontal arrows are
given by the spinorial Lie derivative $(V,\Psi) \mapsto \eL_V \Psi$:
\begin{equation*}
  \begin{CD}
    \gB(\Tilde X) \times \gF(\Tilde X) @>>> \gF(\Tilde X)\\
    @AAA       @VVV\\
    \gB(X) \times \gF(X) @>>> \gF(X)
  \end{CD}
\end{equation*}
We will actually prove something a little more general.  Let $\Psi$ be
any spinor on $\Tilde X$ which restricts to a spinor on $X$ of the
same name.  Let $V$ be a Killing vector on $X$ and let $\Tilde V$ be
the lift to a Killing vector on $\Tilde X$.  We claim that the
spinorial Lie derivative $\eL_{\Tilde V} \Psi$ is a spinor on $\Tilde
X$ which restricts to $\eL_V \Psi$ on $X$.

\begin{description}
\small
\item[\textbf{Proof}]\hspace*{\fill}\linebreak[4]
  Let $\{E_i\}$ be an orthonormal frame for $X$ and let $\{\Tilde
  E_I\} = \{\Tilde E_i=\frac{1}{r} E_i, \Tilde E_r = \d_r\}$ be an
  orthonormal frame for $\Tilde X$.  If $V= V^i E_i$ then the lift
  $\Tilde V = \Tilde V^I \Tilde E_I$ has components: $\Tilde V^r = 0$
  and $\Tilde V^i = r V^i$.  Let us compute the spinorial Lie
  derivative $\eL_{\Tilde V} \Psi$:
  \begin{align*}
    \eL_{\Tilde V} \Psi &= \Tilde\nabla_{\Tilde V} \Psi - \tfrac{1}{4} 
    \Tilde\nabla_I \Tilde V_J \cdot \Gamma^{IJ} \Psi\\
    &= V^i \left(\nabla_i \Psi - \half \Gamma_i\Gamma_r\cdot
    \Psi\right) - \left(\tfrac{1}{4} \nabla_i V_j \Gamma^{ij} \cdot
    \Psi + \half V_j \Gamma^r \Gamma^j\cdot \Psi\right)\\
    &= \nabla_V \Psi - \tfrac{1}{4} \nabla_i V_j \Gamma^{ij} \cdot
    \Psi\\
    &= \eL_V \Psi~.
  \end{align*}
\end{description}

Suppose then that we want to compute the spinorial Lie derivative
$\eL_V \Psi$ of a Killing spinor $\Psi$ on $X$ in the direction of a
Killing vector $V$.  We proceed as follows:
\begin{enumerate}
\item we lift $V$ and $\Psi$ to a Killing vector and a parallel
  spinor, respectively, on the cone $\Tilde X$;
\item compute the spinorial Lie derivative on $\Tilde X$; and
\item we restrict the resulting parallel spinor to a Killing spinor on 
  $X$.
\end{enumerate}
This seemingly circuitous way of computing the spinorial Lie
derivative has the advantage that for the spaces in question the
spinorial Lie derivative on the cone can be computed using elementary
group theory.  Indeed, let $V$ be a Killing vector on $X$ lifted to
$\Tilde X$ and let $\Psi$ be a parallel spinor $\Tilde \nabla \Psi =
0$, then the spinorial Lie derivative $\eL_V$ simplifies to an
infinitesimal orthogonal transformation
\begin{equation}
  \label{eq:LieDerivativeParallel}
  \eL_V \Psi = \Tilde \nabla_V \Psi + \theta(V) \cdot \Psi = \theta(V) 
  \cdot \Psi~.
\end{equation}
In other words, the isometry algebra acts on the spinors as a
subalgebra of the orthogonal algebra in such a way that it preserves
the singlets under the holonomy subalgebra, i.e., the parallel
spinors.

Let us now discuss this briefly case by case.  Of the geometries
listed in Table~\ref{tab:baer}, the only ones which generically have
isometries are the ones possessing Sasakian structures:
Sasaki--Einstein and 3-Sasaki, so we will discuss them in turn.
We will not need the details of what a Sasaki structure is.  The
interested reader can consult \cite{BoGa,BoGa2}.

\subsection{Sasaki--Einstein manifolds}

Bär \cite{Baer} exhibited a one-to-one correspondence between
Sasaki--Einstein structures on a manifold $X$ and a Calabi--Yau
metrics on the cone $\Tilde X$.  For our purposes the main feature of
a Sasaki--Einstein space is the existence of a Killing vector, $S$,
constructed as follows.  Let $\xi$ be the Euler vector on $\Tilde X$.
Since $\Tilde X$ is Kähler, there is a parallel complex structure
$J$.  Let $S \equiv J\xi$.  It is clear that $S$ is orthogonal to
$\xi$ and that it has unit norm.  We claim that it is a Killing
vector.  Indeed,
\begin{equation*}
  \Tilde g(\Tilde \nabla_V S, W) = \Tilde g(J\, \Tilde \nabla_V \xi,
  W) = \Tilde g(J\, V, W) = \omega(V,W)~,
\end{equation*}
where $\omega$ is the Kähler form.  Since this is antisymmetric in
$V,W$, it follows that $S$ is a Killing vector.  Let $\Psi$ be a
parallel spinor in $\Tilde X$. The Lie derivative $\eL_S \Psi$ can be
computed using equation \eqref{eq:LieDerivativeParallel}.  One gets
\begin{equation}
  \label{eq:U(1)}
  \eL_S \Psi = -\tfrac{1}{4} \Tilde\nabla_I S_J \Gamma^{IJ} \cdot
  \Psi = -\tfrac{1}{4} \omega_{IJ} \Gamma^{IJ}\cdot \Psi~.
\end{equation}
We now use the fact that $\tfrac{1}{4} \omega_{IJ}\Gamma^{IJ}$ spans a
very particular $\u_1$ subalgebra of the maximal $\u_n$ subalgebra of
$\so_{2n}$.  Indeed, $\u_n \cong \u_1 \times \su_n$, where $\u_1$ is
the Lie subalgebra generated by the complex structure and
$\su_n\subset\so_{2n}$ is the holonomy algebra of the Calabi--Yau. (We
take $\dim X = 2n-1$.)  To see how this $\u_1$ subalgebra acts on the
parallel spinors, we simply decompose the relevant spinorial
representation of $\so_{2n}$ under $\u_n$ and see how the
$\su_n$ singlets transform under the $\u_1$.

Although it is possible to give a general answer, we will only consider
the two cases of physical interest \cite{AFHS}: $\dim X = 5,7$.
Our results are to be compared with those of Moroianu \cite{Moroianu}
who uses somewhat different methods to obtain similar results.

Consider the case of $\dim X = 5$.  Because $\so_6 \cong \su_4$, the
spin representation is complex and four dimensional.  In fact it is
the $\repre{4}$ of $\su_4$.  The spinors, being real, transform under
the real representation $\real{\repre{4}}$ of
$\repre{4}\oplus\repre{4}^*$.  Under
$\so_6\supset\u_3\cong\u_1\times\su_3$, the real spinor representation
$\real{\repre{4}}$ breaks up as $\real{\repre{4}} \to
\real{\repre{3}_{+1}} \oplus \real{\repre{1}_{-3}}$.  Therefore the
$\su_3$ singlets transform under $\u_1$ as a real two-dimensional
representation $[(+3) \oplus (-3)]$.  In other word, under $\u_1\cong
\so_2$, the parallel spinors transform as a $\repre{2}$.

Now consider the case $\dim X = 7$, so that the cone is a Calabi--Yau
fourfold.  Under $\so_8 \supset \u_1\times\su_4$, the spinor
representation $\repre{8}_s$ breaks up as $\repre{8}_s \to
(\repre{2},\repre{1}) \oplus (\repre{1},\repre{6})$, so that the
parallel spinors again transform as the $\repre{2}$ of
$\u_1\cong\so_2$.

\subsection{3-Sasaki manifolds}

In \cite{Baer} Bär also exhibited a one-to-one correspondence between
3-Sasaki structures on $X$ and hyperkähler metrics on the cone $\Tilde
X$.  On a hyperkähler $\Tilde X$ we have a triplet of parallel complex
structures $I^\alpha$ for $\alpha=1,2,3$ obeying the algebra of
imaginary quaternions:
\begin{equation}
  \label{eq:UnitQuaternions}
  I^\alpha \, I^\beta = - \delta_{\alpha\beta} \1 +
  \epsilon_{\alpha\beta\gamma}\, I^\gamma~.
\end{equation}
Let us define the vector fields $S^\alpha \equiv -\half
I^\alpha\, \xi$, with $\xi$ the Euler vector.  As in the
Sasaki--Einstein case above, each of the $S^\alpha$ is a unit-norm
Killing vector on $\Tilde X$.  Moreover,  because of equation
\eqref{eq:UnitQuaternions}, they obey an $\so_3$ Lie algebra:
\begin{align*}
  [S^\alpha, S^\beta]
  &= \Tilde\nabla_{S^\alpha} S^\beta -
  \Tilde\nabla_{S^\beta} S^\alpha\\
  &= -\half I^\beta \Tilde\nabla_{S^\alpha} \xi + 
  \half I^\alpha \Tilde\nabla_{S^\beta} \xi\\
  &= -\half I^\beta\, S^\alpha + \half I^\alpha\,
  S^\beta\\
  &= \tfrac{1}{4} I^\beta\, I^\alpha\, \xi  - \tfrac{1}{4}
  I^\alpha\, I^\beta\, \xi\\
  &= -\half \epsilon_{\alpha\beta\gamma}\, I^\gamma\, \xi\\
  &= \epsilon_{\alpha\beta\gamma}\, S^\gamma~.
\end{align*}
Now let $\Psi$ be a parallel spinor in $\Tilde X$. Using equation
\eqref{eq:LieDerivativeParallel} we can compute the Lie derivative
$\eL_{S^\alpha} \Psi$:
\begin{equation}
  \label{eq:Sp(1)}
  \eL_{S^\alpha} \Psi = -\tfrac{1}{4} \Tilde\nabla_I
  S^\alpha_J \Gamma^{IJ} \cdot \Psi = \tfrac{1}{8}
  \omega^\alpha_{IJ} \Gamma^{IJ}\cdot \Psi~,
\end{equation}
where $\omega^\alpha$ is the Kähler form associated with the complex
structure $I^\alpha$.  Now we use the fact that the three elements
$\tfrac{1}{8} \omega^\alpha_{IJ}\Gamma^{IJ}$ span an $\sp_1$
subalgebra of $\so_{4n}$ (where $\dim X = 4n - 1$), which is the
centraliser of the holonomy subalgebra $\sp_n \subset \so_{4n}$.
Indeed, $\sp_1 \times \sp_n \subset \so_{4n}$ is a maximal subalgebra.
Just as in the Sasaki--Einstein case, to see how this $\sp_1$
subalgebra acts on the parallel spinors, it is enough to decompose the
relevant spinorial representation of $\so_{4n}$ under $\sp_1 \times
\sp_n$ and see how the $\sp_n$ singlets transform under the $\sp_1$.

Again, although it is possible to give a more general answer, we limit
ourselves to the case of interest: $n=2$, so that $\dim X = 7$.  In
this case, the spinor representation $\repre{8}_s$ of $\so_8$ breaks
up under $\sp_1\times \sp_2 \subset \so_8$ as $\repre{8}_s \to
(\repre{3},\repre{1}) \oplus (\repre{1},\repre{5})$.  Thus we see that
the three parallel spinors transform as the $\repre{3}$ of $\sp_1
\cong \so_3$.

\subsection{The symmetry superalgebras}

Finally we can put it all together and identify the symmetry
superalgebras.  We are interested in geometries of the form $\ads_4
\times X_7$, $\ads_5 \times X_5$ and $\ads_7 \times X_4$.  We have
several choices for $X$, given by Table~\ref{tab:baer}, whose
\emph{generic} isometry algebras $\gB(X)$ are given by the second
factor in the $\gB$ column of
Table~\ref{tab:superalgebras}.\footnote{We say generic because some of
  these manifolds will have larger isometry algebras.  However it is
  possible to prove (see, e.g., \cite{Moroianu} for some results in
  this direction) that the larger isometry algebra is a product of the
  generic isometry algebra and a compact Lie algebra which acts
  trivially on the Killing spinors.  This is nothing but the geometric 
  manifestation of the
  Coleman--Mandula--Haag--{\L}opusza\'nsky--Sohnius theorem.}
The first factor in that column corresponds to the isometry algebra
$\so_{p+1,2}$ of $\ads_{p+2}$.  For each of the geometries in
Table~\ref{tab:superalgebras} we have determined the representation
$\gF$ of the isometry algebra that the Killing spinors are in.  The
$X$ part follows from the above considerations, whether the $\ads$
part is well-known and will not be rederived here.  Finally, equipped
with the Lie algebra $\gB$ and the representation $\gF$ we look up in
Nahm's classification \cite{Nahm} of superconformal algebras and we
see that in each case there is a unique superconformal algebra with
that data, which is the superconformal algebra $\gS$ listed in the
Table.

As a final comment, let us remark that whereas the method presented in
this note is quite general, group theory would not be enough to
determine the precise symmetry superalgebra treated in
\cite{GMT1,PKT}.  That superalgebra has a free parameter $\alpha$
taking values in the unit interval, which cannot be determined simply
from a knowledge of $\gB$ and $\gF$.  In this case one must compute
the Killing spinors explicitly as was done in \cite{GMT1,PKT}, and
determine the value of $\alpha$ by looking at the symmetric bilinear
$\gF \otimes \gF \to \gB$.

\section*{Acknowledgements}
It is a pleasure to thank Andrei Moroianu, Bill Spence, Sonia Stanciu
and Stefan Vandoren for useful discussions, and Paul Townsend for the
encouragement to write this thing up.  This work was partially funded
by the EPSRC, whom I would like to thank for their support.

%

\begin{thebibliography}{10}

\bibitem{AFHS}
BS~Acharya, JM~Figueroa-O'Farrill, CM~Hull, and B~Spence, \emph{Branes at
  conical singularities and their dual field theories}, Adv. Theor. Math. Phys.
  in press, \texttt{hep-th/9808014}.

\bibitem{BoGa}
CP~Boyer and K~Galicki, \emph{3-{Sasakian} manifolds}, \texttt{hep-th/9810250}.

\bibitem{BoGa2}
\bysame, \emph{On {S}asakian--{E}instein geometry}, \texttt{math.DG/9811098}.

\bibitem{Baer}
C~Bär, \emph{Real {K}illing spinors and holonomy}, Comm. Math. Phys.
  \textbf{154} (1993), 509--521.

\bibitem{CheegerEbin}
J~Cheeger and D~Ebin, \emph{Comparison theorems in {R}iemannian {G}eometry},
  North-Holland, Amsterdam, 1975.

\bibitem{Gallot}
S~Gallot, \emph{Equations différentielles caractéristiques de la sphère}, Ann.
  Sci. École Norm. Sup. \textbf{12} (1979), 235--267.

\bibitem{GMT2}
JP~Gauntlett, RC~Myers, and PK~Townsend, \emph{Black holes of ${D}=5$
  supergravity}, Class. \& Quantum Grav. \textbf{16} (1999), 1--21,
  \texttt{hep-th/9810204}.

\bibitem{GMT1}
\bysame, \emph{Supersymmetry of rotating branes}, Phys. Rev. \textbf{D59}
  (1999), 025001, \texttt{hep-th/9809065}.

\bibitem{GR2}
GW~Gibbons and P~Rychenkova, \emph{Cones, tri-sasakian structures and
  superconformal invariance}, Phys. Lett. \textbf{B443} (1998), 138--142,
  \texttt{hep-th/9809158}.

\bibitem{KlebanovWitten}
IR~Klebanov and E~Witten, \emph{Superconformal field theory on threebranes at a
  {C}alabi--{Y}au singularity}, Nucl. Phys. \textbf{B536} (1998), 199--218,
  \texttt{hep-th/9807080}.

\bibitem{Kosmann}
Y~Kosmann, \emph{Dérivées de {L}ie des spineurs}, Annali di Mat. Pura Appl.
  (IV) \textbf{91} (1972), 317--395.

\bibitem{Malda}
JM~Maldacena, \emph{The large {$N$} limit of superconformal field theories and
  supergravity}, Adv. Theor. Math. Phys. \textbf{2} (1998), 231--252,
  \texttt{hep-th/9711200}.

\bibitem{Moroianu}
A~Moroianu, \emph{On the infinitesimal isometries of manifolds with {K}illing
  spinors}, Humboldt Universität preprint.

\bibitem{Nahm}
W~Nahm, \emph{Supersymmetries and their representations}, Nucl. Phys.
  \textbf{B135} (1978), 149--166.

\bibitem{Obata}
M~Obata, \emph{Certain conditions for a riemannian manifold to be isometric
  with a sphere}, J. Math. Soc. Japan \textbf{14} (1962), 333--340.

\bibitem{Spindel}
P~Spindel, \emph{Gravity before supergravity}, Supersymmetry (K~Dietz, R~Flume,
  G~von Gehlen, and V~Rittenberg, eds.), NATO Series B, vol. 125, Plenum (New
  York), 1984, pp.~455--533.

\bibitem{PKT}
PK~Townsend, \emph{Killing spinors, supersymmetries and rotating intersecting
  branes}, \texttt{hep-th/9901102}.

\bibitem{Wang}
MY~Wang, \emph{Parallel spinors and parallel forms}, Ann. Global Anal. Geom.
  \textbf{7} (1989), no.~1, 59--68.

\end{thebibliography}
%

\providecommand{\bysame}{\leavevmode\hbox to3em{\hrulefill}\thinspace}

\end{document}